\PassOptionsToPackage{prologue,dvipsnames,table}{xcolor}
\documentclass[acmsmall,screen]{acmart}  

\usepackage{amsmath}
\usepackage[frozencache=true,cachedir=minted-cache]{minted}
\usepackage{multirow}
\usepackage[font=small]{caption}
\usepackage{subcaption}
\usepackage{hyperref}
\usepackage[T1]{fontenc}

\usepackage{listings}
\usepackage[most]{tcolorbox}
\usepackage{fancyvrb}
\usepackage{changepage}
\usepackage{csvsimple}  %
\usepackage{booktabs}   %
\usepackage{array}      %
\usepackage{multirow}
\usepackage{enumitem}
\usepackage{ragged2e}

\usepackage{algorithm}
\usepackage{algorithmicx}
\usepackage{algpseudocode}

\usepackage{rotating}

\lstnewenvironment{run}[1]{}{}

\NewTColorBox{mybox}{ O{red} m d"" !O{} }{
    enhanced,
    colframe=#1!75!black,
    colback=#1!5!white,
    fonttitle=\bfseries,#2,
    beforeafter skip=0.25em,
    top=0em,
    bottom=0em,
    left=0.5em,
    right=0.25em,
    boxrule=0.25pt
}

\newcommand{\prompt}[1]{\texttt{\textbf{(ChatDBG) #1}}}

\newenvironment{trace}[0]{
    \scriptsize
    \begin{mybox}[gray]{}
}{
    \end{mybox}
}

\newenvironment{user}[1]{
    \prompt{#1}
}{
}

\newenvironment{user_special}[1]{
    \texttt{\textbf{#1}}
}{
}

\newenvironment{assistant}{
    \begin{mybox}[blue]{}
}{
    \end{mybox}
}

\newenvironment{function}[1]{
    \begin{mybox}[yellow]{}[colbacktitle=yellow!50!white,coltitle=black]
    \prompt{#1}
    \vspace*{-0.25em}
}{
    \end{mybox}
}

\newenvironment{prose}[0]{
    \vspace*{0.25em}
    \setlength{\parskip}{0.5em}
    \sffamily
    \raggedright
    \begin{adjustwidth}{.5em}{1em}      
}{
    \end{adjustwidth}
    \vspace*{0.2em}
}

\DefineVerbatimEnvironment{Highlighting}{Verbatim}{commandchars=\\\{\}}
\newenvironment{Shaded}{}{}

\newcommand{\ControlFlowTok}[1]{#1}

\newcommand{\KeywordTok}[1]{#1}
\newcommand{\NormalTok}[1]{#1}
\newcommand{\OperatorTok}[1]{#1}

\newcommand{\StringTok}[1]{#1}

\renewcommand{\subparagraph}[1]{\textbf{#1}}

\providecommand{\tightlist}{%
  \setlength{\itemsep}{0pt}\setlength{\parskip}{0pt}}

\newcommand{\config}[1]{\textbf{#1}}

\newcommand{\systemname}{\textsc{ChatDBG}}

\title{\systemname{}: Augmenting Debugging with Large Language Models}



\setcopyright{cc}
\setcctype{by-nd}
\acmDOI{10.1145/3729355}
\acmYear{2025}
\acmJournal{PACMSE}
\acmVolume{2}
\acmNumber{FSE}
\acmArticle{FSE085}
\acmMonth{7}
\received{2024-09-12}
\received[accepted]{2025-04-01}

\author{Kyla H. Levin}
\authornote{Equal contribution.}
\orcid{0009-0005-2533-7499}
\affiliation{
    \institution{University of Massachusetts Amherst}
    \city{Amherst}
    \state{MA}
    \country{USA}
}
\email{khlevin@cs.umass.edu}

\author{Nicolas van Kempen}
\authornotemark[1]
\orcid{0000-0002-1708-0073}
\affiliation{
    \institution{University of Massachusetts Amherst}
    \city{Amherst}
    \state{MA}
    \country{USA}
}
\email{nvankempen@cs.umass.edu}

\author{Emery D. Berger}
\authornote{Work done at the University of Massachusetts Amherst.}
\orcid{0000-0002-3222-3271}
\affiliation{
    \institution{University of Massachusetts Amherst}
    \city{Amherst}
    \state{MA}
    \country{USA}
}
\affiliation{
    \institution{Amazon Web Services}
    \city{Seattle}
    \state{WA}
    \country{USA}
}
\email{emery@cs.umass.edu}

\author{Stephen N. Freund}
\orcid{0009-0000-6992-199X}
\affiliation{
    \institution{Williams College}
    \city{Williamstown}
    \state{MA}
    \country{USA}
}
\email{freund@cs.williams.edu}

\begin{document}

\begin{abstract}
    Debugging is a critical but challenging task for programmers. This paper
proposes \systemname{}, an AI-powered debugging
assistant. \systemname{} integrates large language models (LLMs) to
significantly enhance the capabilities and user-friendliness of
conventional debuggers. \systemname{} lets programmers engage in a
collaborative dialogue with the debugger, allowing them to pose
complex questions about program state, perform root cause analysis for
crashes or assertion failures, and explore open-ended queries like
`\texttt{why is x null?}'. To handle these queries, \systemname{}
grants the LLM autonomy to \emph{take the wheel}: it can act as an independent agent capable of querying and
controlling the debugger to navigate through stacks and inspect program
state. It then reports its findings and yields back control to the
programmer. By leveraging the real-world knowledge
embedded in LLMs, \systemname{} can diagnose issues identifiable only through
the use of domain-specific reasoning. Our \systemname{} prototype integrates with standard
debuggers including \texttt{LLDB} and \texttt{GDB}
for native code and \texttt{Pdb} for Python. Our evaluation across a
diverse set of code, including C/C++ code with known bugs and a suite
of Python code including standalone scripts and Jupyter notebooks,
demonstrates that \systemname{} can successfully analyze root causes,
explain bugs, and generate accurate fixes for a wide range of
real-world errors. For the Python programs, a single query led to an
actionable bug fix 67\% of the time; one additional follow-up query
increased the success rate to 85\%. \systemname{} has seen rapid
uptake; it has already been downloaded more than 75,000 times.

\end{abstract}

\maketitle

\section{Introduction}
\label{sec:introduction}

Debuggers help programmers identify and
fix bugs by letting them investigate program state and navigate
program execution. Debuggers for mainstream languages, including
GDB~\cite{stallman-debugging} and LLDB~\cite{lldb} (for C,
C++, and Rust), JDB (for Java), Pdb (for Python), and the Chrome or Firefox
debuggers (for JavaScript), generally provide the same functionality.
In particular, most debuggers support
  observing program execution via \emph{tracing} and reporting when a
    program reaches a given line or function of source
    code;
interrupting execution and returning control to the debugger when
    the program reaches a given line or function via
    \emph{breakpoints}, when a particular condition is true via
    \emph{conditional breakpoints}, or when a variable changes via
    \emph{watchpoints} (a.k.a. \emph{data breakpoints});
inspecting local
    variables, globals, heap objects, and \emph{backtraces} of the call stack; and
resuming program execution line-by-line (\emph{single-step}) or at
    the granularity of function calls.

Debuggers can be helpful, but finding and fixing
software defects remains a deeply challenging and time-consuming
task~\cite{ZellerDebugging, eisenstadt1993tales, DBLP:conf/esem/LaymanDNSDV13}.
Programmers must still reason about program
behavior to ascertain what went wrong. They must formulate and test hypotheses
about program execution,
they must read and understand code they may have not written,
and they must pore over potentially voluminous information. Such information
includes lengthy executions, large amounts of program data, and
many stack frames that potentially span multiple threads.

This paper introduces the \textbf{\systemname{}} AI-powered
debugger assistant. \systemname{} integrates into and significantly
extends the functionality of standard debuggers.  
\systemname{} builds
on the insight that large language models (LLMs), such as OpenAI's
GPT-4~\cite{openai2024gpt4technicalreport}, enable a
debugger to leverage insights and intuition from
the vast real-world knowledge embedded in LLMs.
This knowledge enables \systemname{} to fix classes of issues that depend on
logical thinking and domain-specific reasoning
beyond the ability to write and debug programs.
For example, Figure \ref{fig:bootstrap} illustrates the use of \systemname{} to debug a program
by leveraging a knowledge of statistics that cannot be gleaned from the program itself.

A debugger integrated with \systemname{} continues to provide its
full range of functionality but also lets programmers engage in
debugging dialogs where they can ask high-level questions like
`\texttt{why is x null here?}' or `\texttt{why isn't this value what I
  expected?}'. The question can be as simple as `\texttt{why?}' if
a program has crashed or failed an assertion.
To answer such queries, \systemname{} orchestrates a conversation with an LLM.  A key
feature of \systemname{} is that it grants autonomy to the LLM to
``take the wheel'' and act as an independent
agent~\cite{DBLP:journals/ker/WooldridgeJ95, DBLP:conf/iclr/GurFHSMEF24}
while answering the programmer's queries. Specifically, the LLM issues ``function
calls''~\cite{OpenAIFunctions} to run commands in the underlying
debugger to investigate program state, execute code, or obtain source
code.  The results of those calls are sent back to the LLM to use
in constructing its response.  After answering a query, control is
returned to the programmer, who may then enter additional commands or chat
messages.

Our prototype of \systemname{} integrates
into three widely used debuggers: GDB, LLDB, and Pdb. Our
evaluation presents a range of case studies demonstrating that
\systemname{} improves significantly on existing debuggers.  On a
suite of unpublished Python scripts and Jupyter notebooks written by
undergraduate students, one or two queries is sufficient for
\systemname{} to properly diagnose and fix defects 85\% of the time,
typically at a cost well under \$0.20 USD.  \systemname{} is also
effective at identifying causes and providing fixes for a range of
real-world bugs in C/C++ code.

This paper makes the following contributions:
\begin{itemize}
  \item It introduces
\systemname{}, an AI-powered debugger assistant that enables large language models to ``take the wheel''
and control the debugger via agentic reasoning.
\item It describes
the implementation of our \systemname{} prototype.
\item It presents an evaluation of
\systemname{} that demonstrates its significant advantages over existing
debugger functionality.
\end{itemize}

Our evaluation shows that \systemname{} is broadly applicable to many domains and programming
languages, and we expect it to be particularly useful for
novice programmers, who often lack the experience to effectively use debuggers.
\systemname{} is also useful for experienced
programmers, who can augment debugging sessions with \systemname{}'s reasoning capabilities
in a conversational and interactive way.

\begin{figure}[tp!]
  \textbf{Source code for bootstrap.py}
  \begin{center}
\begin{minted}{python}
from datascience import *
from ds101 import *

def make_marble_sample():
    table = Table().read_table('marble-sample.csv')
    return table.column('color')

def proportion_blue(sample):
    return sample

def resampled_stats(observed_marbles, num_trials):
    stats = bootstrap_statistic(observed_marbles,
                                proportion_blue,
                                num_trials)
    assert len(stats) == num_trials
    return stats

observed_marbles = make_marble_sample()
stats = resampled_stats(observed_marbles, 5)

assert np.isclose(np.mean(stats), 0.7)
\end{minted}
  \end{center}
  \caption{\label{fig:bootstrap-src} \textbf{An example program
      containing several bugs (\textsection\ref{sec:overview}).}  It is
    supposed to create an array of marble colors, compute the proportions of blue
    marbles in resamples of that array, and assert that their
    mean is about 0.7, the proportion for the array.}
  \Description{A Python code snippet.}
\end{figure}

\begin{figure}[tp!]
  \begin{minipage}{\columnwidth}
    \input{figs/bootstrap1-trace.tex}   
  \end{minipage}
  \caption{\label{fig:bootstrap} \textbf{A debugging session with
      \systemname{} (\textsection\ref{sec:overview}).} For brevity, we elide repetitive or unimportant parts. \systemname{} catches the assertion failure on
    line 15 and prompts the user to enter commands (\textbf{bold} and highlighted in \colorbox{gray!10}{gray}).
    Commands issued by \systemname{} when it takes the wheel are
    highlighted in \colorbox{yellow!10}{yellow}, and its response is
    highlighted in \colorbox{blue!10}{blue}.}
    \Description{Text snippet of a \systemname{} debugging session.}
\end{figure}

\section{Overview}
\label{sec:overview}

\renewcommand{\t}[1]{\mbox{\texttt{#1}}}
\newcommand{\inlinecode}[1]{\lstinline[basicstyle=\tt]{#1}}

This section illustrates \systemname{}'s features and ability to
assist in debugging the program in Figure~\ref{fig:bootstrap-src}.
That program is a distillation of real errors encountered by students in
an introductory data science lab.  It creates an array
\inlinecode{observed_marbles} representing the colors of marbles (red
or blue) in a sample stored in a file.  It then calls
\inlinecode{bootstrap_statistic} to create same-sized resamples of
that array. That function computes a statistic for each resample and
returns an array of those statistics. In this case, the statistic is
\inlinecode{proportion_blue}, the proportion of blue marbles. Given a
sufficiently large number of trials, the mean of the resamples' statistics should
be close to 0.7, the proportion of blue marbles in the original
sample~\cite{e89fac9c-03d7-3e22-aa30-08f5596f8fce}.

The program fails the assertion in \texttt{resampled\_stats}, and
Figure~\ref{fig:bootstrap} illustrates a debugging session.  To
try to figure out what went wrong, the user issues the Pdb command
\inlinecode{p num_trials} to view the value of that variable. Continuing
debugging with existing tools would likely involve issuing additional
commands, examining data files, source code, and examining library
documentation.
With \systemname{}, the user instead starts a dialog with the
debugger, asking \texttt{why doesn't stats have 5 elements?}  While
constructing the answer (in \colorbox{blue!10}{blue}), the LLM
\emph{takes the wheel} and directly issues debugger commands
(\colorbox{yellow!10}{yellow}). These include standard Pdb commands
and a \systemname{}-specific \inlinecode{info} command for accessing the source
code and docstrings for any user-written code, as well as the docstrings for
library code (which we assume is correct and not the root cause of
errors).

\begin{figure}
    \input{figs/bootstrap3-trace.tex}
    \caption{\label{fig:bootstrap2}\textbf{A debugging session
        demonstrating how \systemname{} incorporates real-world
        knowledge (\textsection\ref{sec:overview}).} After correcting %
      \inlinecode{proportion_blue} in Figure~\ref{fig:bootstrap-src},
      the program fails on line 21 because the mean proportion of blue
      marbles in the resamples is not the expected value.
      \systemname{} identifies high variance resulting from the small
      number of trials as the root cause.}
      \Description{Text snippet of a \systemname{} debugging session.}
\end{figure}

\systemname{} identifies and corrects the root cause: \inlinecode{proportion_blue}
incorrectly computes the desired statistic.
When \systemname{} cannot identify the root cause, it suggests further
debugging steps and control is returned to the user, who may continue
the chat, issue further debugger commands, or both.
Figure~\ref{fig:bootstrap2} illustrates this scenario, where a version
of \t{bootstrap.py} with the corrected \inlinecode{proportion_blue}
function fails the assertion on line 21. 

The user asks why the mean
of \inlinecode{stats} is not close to 0.7, and \systemname's initial
response suggests examining whether 0.7 is the appropriate expected
value.  To verify this, the user then computes the proportion of blue marbles with a
debugger command and tells \systemname{} that 0.7 is indeed the correct
value.  In its follow-up, \systemname{} points to the low number of trials
(five) as the issue.  The LLM drew this correct
conclusion without seeing any discussion of trial size or variance in
any program state, code, or documentation encountered during
the chat. A powerful aspect of \systemname{} is its ability to exploit
real-world knowledge in its analyses (here, the fact that
bootstrapping depends on large numbers of resamples) without specific
instruction or user intervention.

\section{Related Work}
\label{sec:relatedwork}

\begin{table}[!t]
  \centering
  \footnotesize
\caption{\textbf{Debugger features and their dates of introduction (\textsection\ref{sec:relatedwork}).}
Most key features have been around for decades. By integrating into
modern debuggers (GDB, LLDB, and Pdb), \systemname{} inherits
all of their features while significantly extending them with
functionality to explain bugs and their root causes, propose fixes,
and answer arbitrary natural-language queries over program
state. (An asterisk or \emph{year} in italics means the feature is
limited in functionality, performance, or depends on specific
hardware support.)
\label{tab:debugger-features}}
\begin{tabular}{@{}lp{0.18in}p{0.18in}p{0.18in}p{0.18in}p{0.18in}p{0.18in}p{0.18in}p{0.18in}p{0.18in}p{0.18in}p{0.18in}l@{}}
  \toprule
  \textbf{System and Date} 
  & \rotatebox{45}{\textbf{Single Step}} 
   & \rotatebox{45}{\textbf{Stack Navigation}} 
   & \rotatebox{45}{\textbf{Breakpoints (BPs)}} 
   & \rotatebox{45}{\textbf{Conditional BPs}}
   & \rotatebox{45}{\textbf{Source Level}} 
   & \rotatebox{45}{\textbf{Trace}} 
   & \rotatebox{45}{\textbf{Display State}} 
   & \rotatebox{45}{\textbf{Eval. Code}} 
   & \rotatebox{45}{\textbf{Watchpoints}} 
   & \rotatebox{45}{\textbf{Explain Bugs}} 
   & \rotatebox{45}{\textbf{Propose Fixes}} 
   & \rotatebox{45}{\textbf{Open Queries}} \\
\midrule
DDT~\cite{kotok1961dec}, 1961 & \checkmark & \checkmark & \checkmark & &  &  &  &  &  &  & \\
EXDAMS~\cite{DBLP:conf/afips/Balzer69}, 1969 & \checkmark & \checkmark & \checkmark & &  &  &  &  & \checkmark & & & \\
Mesa~\cite{mesa-debugger}, 1979 & \checkmark & \checkmark & \checkmark & \checkmark* & \checkmark & \checkmark & \checkmark & \checkmark &  &  & & \\
Dbx~\cite{DBLP:conf/usenix/Linton90}, 1981 & \checkmark & \checkmark & \checkmark & 1990 & \checkmark & \checkmark & \checkmark & \checkmark & \checkmark &  & & \\
GDB~\cite{stallman-debugging}, 1986 & \checkmark & \checkmark & \checkmark & \checkmark & \checkmark & \checkmark & \checkmark & \checkmark & 1991 & & \\
Pdb, 1992 & \checkmark & \checkmark & \checkmark & \checkmark & \checkmark & \checkmark & \checkmark & \checkmark & \checkmark &  & & \\ %
LLDB, 2010 & \checkmark & \checkmark & \checkmark & \checkmark & \checkmark & \checkmark & \checkmark & \checkmark & \checkmark &  & & \\
\midrule
\textbf{\systemname{}}, 2023 & \checkmark & \checkmark & \checkmark & \checkmark & \checkmark & \checkmark & \checkmark & \checkmark & \checkmark &  \checkmark & \checkmark & \checkmark \\
\bottomrule
\end{tabular}
\end{table}

Table~\ref{tab:debugger-features} presents an overview of previous
interactive debuggers, together with their features.
The first interactive debugger, DDT, introduced
breakpoints, single-stepping, and stack navigation in
1961~\cite{kotok1961dec}. By 1979, the Mesa debugger had most key
features of modern debuggers, including source-level debugging,
conditional breakpoints, tracing, and the ability to display run-time
state and evaluate code~\cite{mesa-debugger}. Arbitrary conditional
breakpoints date back at least to 1990 with
Dbx~\cite{DBLP:conf/usenix/Linton90}. Watchpoints were introduced by
1991 and have been in GDB since version 3.93.

In other work, Ko and Myers present Whyline, an interactive, trace-based debugger
that lets programmers select from a range of queries and identifies
(via static and dynamic analysis) a timeline that answers the
query~\cite{DBLP:journals/tosem/KoM10}. Programmers can only select
from those queries presented by Whyline as options.  In contrast,
\systemname{} permits programmers to pose arbitrary queries that it
answers via a dialog with an LLM. Whyline's use of traces gives it the
ability to answer questions that might not be straightforward to
answer with the current program state but limits its applicability to
relatively short-lived executions.

The goal of \emph{program slicing}, introduced by Weiser in
1981~\cite{DBLP:conf/icse/Weiser81}, is to produce a shorter version
of a program limited to the source code that could have led to an
error. Program slicing has been extensively studied; Weiser's paper
has been cited over 5,000 times to date. As Section~\ref{sec:slice-command}
describes, \systemname{} performs backwards slicing to collect code spread
across code cells to facilitate debugging of Jupyter notebooks.

\emph{Fault localization} seeks to identify the likely location of a defect's
root cause.
Several prior studies have investigated the use of machine learning and LLMs for fault localization.
Some of the studied techniques apply machine learning to source code features, coverage data, or other static code features 
to predict faulty lines of code, but they do not utilize dynamic state and run-time information.
DeepFL~\cite{DBLP:conf/issta/LiLZZ19}, Grace~\cite{DBLP:conf/sigsoft/LouZDLSHZZ21}
and DeepRL4FL~\cite{DBLP:conf/icse/Li0N21a} are examples of such systems.
Similarly, LLMAO~\cite{DBLP:conf/icse/YangGMH24} employs LLMs to provide
suspiciousness scores for each line of code in a given program, but only 
provides access to the source code.
AutoFL~\cite{DBLP:journals/pacmse/KangAY24} also utilizes an LLM
and enables it to statically retrieve source code and coverage information about the program via function calls.  
However, the system requires a failing test case as input and does not employ run-time state information.

\systemname{} improves upon these systems by providing an LLM with access to
run-time program state and the ability to take control of the underlying debugger. 
Both features enhance the LLM's ability to provide more accurate and informative feedback to the user.
We also note that other fault localization techniques can be used in tandem with \systemname{}
to improve results, as suggested by Section~\ref{sec:python}'s utilization of backwards slicing to
identify code relevant to a bug in Python notebooks and Section~\ref{sec:native}'s utilization of
AddressSanitizer~\cite{DBLP:conf/usenix/SerebryanyBPV12} to provide a better starting point for
diagnosing and fixing memory errors in native code.

\emph{Automated program repair} is another active area of software
engineering research~\cite{DBLP:journals/cacm/GouesPR19}.  Systems for automatic program repair attempt to 
generate source-level program patches that prevent a program from
failing. \systemname{} performs best-effort automated
program repair by requesting that the LLM propose code fixes as part of its response,
ultimately letting the programmer drive code changes using these suggestions.
Previous research has shown that automated program repair hints can provide
significant help in the debugging process and suggests that the benefits of correct advice
outweigh the risk of deceptive ones~\cite{DBLP:conf/icse/EladawyGB24}.

\subsection{Concurrent Work}

Several approaches developed concurrently with \systemname{} have
also integrated LLMs into automatic program repair or fault
localization techniques to enhance the debugging
process. Robin~\cite{DBLP:conf/vl/BajpaiCBACGPRS24} is a chat-based
debugging assistant designed to help users diagnose errors more quickly.
Both Robin and \systemname{} provide a limited program context to the LLM at
the beginning of a conversation. However, Robin has no
direct access to any additional context about the program and
execution state; the user must manually retrieve and provide these items.
Robin's functionality is therefore roughly equivalent to the \config{Enriched Stack}
configuration of \systemname{}, detailed in
Section~\ref{sec:python}. As Figure~\ref{fig:summary} shows, \systemname{}
achieves a
nearly two-fold increase improvement in diagnosing errors versus the
\config{Enriched Stack} configuration. \systemname{}'s effectiveness generally increases further
with targeted questions and
follow-up discussions with the user.

AutoCodeRover~\cite{10.1145/3650212.3680384} and
SWE-agent~\cite{DBLP:journals/corr/abs-2405-15793} are complementary
approaches that focus on fault localization and automatic repair,
relying exclusively on issue descriptions and source code.
\systemname{} additionally leverages run-time information to identify
root causes and propose fixes. Section~\ref{sec:evaluation} demonstrates the strength of this approach over relying solely on static information.
Both AutoCodeRover and SWE-agent perform an evaluation using SWE-bench~\cite{DBLP:conf/iclr/JimenezYWYPPN24},
which was created to evaluate the efficacy of such static tools; unfortunately, 
this benchmark suite is not applicable to \systemname{}
due to its extensive usage of run-time information.

RepairAgent~\cite{DBLP:journals/corr/abs-2403-17134} and
AutoSD~\cite{DBLP:journals/ese/KangCYL25} are tools that employ LLMs in
specific workflows that mimic standard debugging strategies in an attempt to repair
pre-identified bugs.
While successful in some settings, both tools rely on the user providing a failing test case and the precise location of the
bug. By contrast, \systemname{} does not require this information. \systemname{} also enables a more flexible workflow that permits
collaboration with the developer in addition to seamless integration into the standard debugging process.

\section{Implementation}
\label{sec:implementation}

\subsection{Using \systemname{}: Preliminaries}

\systemname{} integrates with existing debuggers as either a plug-in
or a direct extension.
Our primary focus %
to date has
been an extension to Pdb, which supports both non-interactive Python
scripts and interactive sessions in IPython or Jupyter notebooks, and
a plug-in for LLDB to support C/C++ code.  A subset of features
has been ported to GDB and WinDBG.

Configuration for Python is minimal and limited to
the installation of the \t{chatdbg} package with the standard package
installer, plus one optional shell script command to add it as an
extension to IPython.  \systemname{} extends either the standard
\t{pdb.Pdb} debugger or IPython's implementation of Pdb, depending on
how it is run.  Configuration for LLDB and other C/C++ debuggers is
similarly straightforward.
LLDB can be installed through standard package managers if it is not
already present, and the \systemname{} plug-in is installed via a single
shell command.
Since \systemname{} leverages OpenAI's LLMs, the user must also set an
environment variable to a valid OpenAI API key within their system's
configuration settings.

\subsection{Debugging a Target Program}

For Python, debugging with \systemname{} begins by running %
\t{chatdbg} on the target program.  No special preparation of the
target is needed; Python's managed run time ensures that debugging
information and source code is always available.  Debugging is
supported in IPython interactive sessions or Jupyter notebooks via the
standard command-line flag \inlinecode{--pdb} or the Jupyter magic
command \t{\%pdb}, respectively.  Control drops into the debugger when
an exception occurs.

For C and C++, debugging begins by running \t{lldb}
on the target program.  The target program must be an unstripped
executable generated with the \texttt{-g} compiler flag, which ensures
the availability of DWARF debug information that describes the memory
layout and maps the program's machine code back to the original source
code.  That information is essential for the effective debugging of
unmanaged code.

\systemname{} also handles native code generated for other languages
but may require additional steps.  For example, to debug a Rust target
program, the \texttt{Cargo.toml} file must list \systemname{} as a
dependency and the \texttt{main} function must be annotated with
\texttt{\#[chatdbg::main]} to ensure that error messages are visible
to \systemname{} through a log file.

\newcommand{\circled}[1]{\raisebox{.5pt}{\textcircled{\raisebox{-.9pt} {\bf #1}}}}
\newcommand{\circlemark}[2]{\textcolor{#2}{\circled{#1}}}
\newcommand{\paddedcirclemark}[2]{\hfill\textcolor{#2}{\circled{#1}}~~~~~}

\begin{figure*}[t!]
  \begin{tabular}{c@{\quad~~}@{\quad}c}
    \begin{minipage}{0.5\columnwidth}
      \includegraphics[width=\columnwidth]{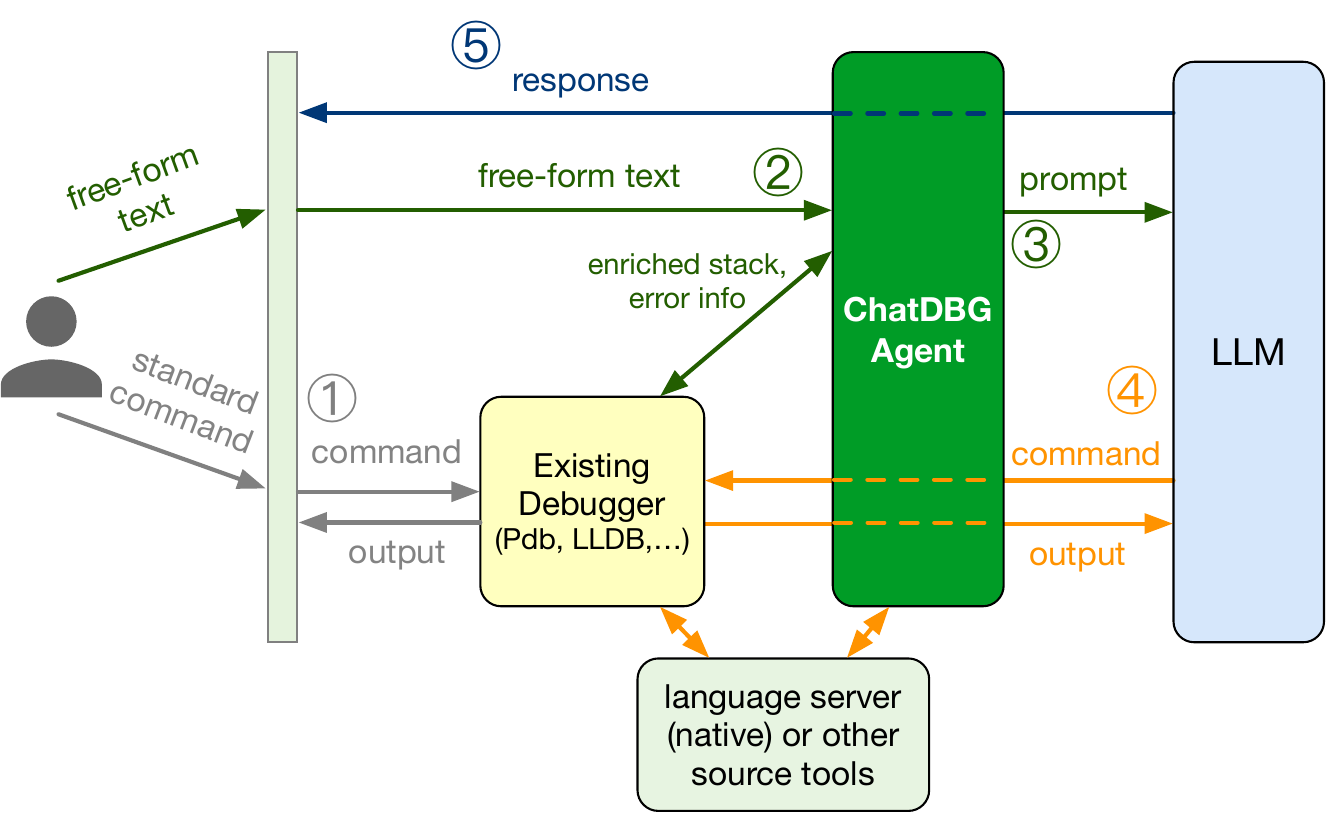} \\
      \footnotesize
      \begin{description}[labelindent=0em,leftmargin=2em]
        \item[\circlemark{1}{gray}] Standard commands are handled
        by the existing debugger.
        \item[\circlemark{2}{ForestGreen}] \systemname{} converts free-form text
        into a suitable prompt.
        \item[\circlemark{3}{ForestGreen}] \systemname{} sends the
        prompt.
        \item[\circlemark{4}{Orange}] The LLM takes the wheel and directly issues
        commands to the underlying debugger.  This step may involve consulting other tools, such as a language server for native code.
        \item[\circlemark{5}{Blue}] The LLM responds to the prompt.
      \end{description}
    \end{minipage}
    &
    \begin{minipage}{0.95\columnwidth}
      \input{algorithm}
    \end{minipage}
  \end{tabular}
  \caption{\label{fig:arch} \textbf{\systemname{} architecture and
  command processing algorithm (\textsection\ref{section:architecture}).}}
  \Description{A graph showing the architecture of \systemname{}, along with a short algorithm summary.
    Users may directly issue standard debugger commands, that are handled by the existing debugger.
    They may also use free-form text, that will be interpreted by \systemname{}. \systemname{} will act
    as an agent, and facilitate conversation between the LLM and the underlying debugger. The LLM will issue
    commands to retrieve context about source code or runtime state before issuing its final response
    and/or recommendation.
  }
\end{figure*}

\subsection{\systemname{} Architecture Overview}
\label{section:architecture}

\systemname{} orchestrates communication between the user, the
debugger, and the LLM, as shown in the architecture diagram in Figure~\ref{fig:arch}.  The
operations in the command loop pseudocode map naturally onto debugger APIs and onto
LLM APIs supporting completion and function
calls~\cite{OpenAIFunctions}.  \systemname{} currently utilizes
OpenAI's API~\cite{OpenAIApi} and GPT-4 models.  We provide a brief overview of the \systemname{} architecture
and then elaborate on
the most salient technical innovations below.

\circlemark{1}{gray} \systemname{} dispatches standard
commands, such as \inlinecode{p num_trials} in
Figure~\ref{fig:bootstrap}, directly to the underlying debugger (lines
3-7).  It also preserves those commands and their output in the
history variable for later communication to the LLM.
\circlemark{2}{ForestGreen} Any other text entered by the user, such
as `\texttt{why doesn't stats have 5 elements?}', is directed to
\systemname{}, which creates a prompt to send to the LLM.  If this is
the start of a chat, \systemname{} bundles basic instructions,
information from the debugger about the current stack and error, program inputs,
history of user commands, and the text together in an \emph{initial
prompt} (lines 9-12).  Otherwise, \systemname{} bundles only the
history since the last chat step and text (line 14).  The
\textsc{MakePrompt} function concatenates the prompt components
into a string, respecting any length limits set by the LLM by
selectively truncating parts as needed.

\circlemark{3}{ForestGreen} \systemname{} then sends the prompt to the
LLM and processes the response stream, which includes both
\circlemark{4}{Orange} requests to run debugger commands (lines 19-22)
and \circlemark{5}{Blue} prose for the user (line 23).  In
Figure~\ref{fig:bootstrap}, \systemname{} runs four debugging
commands, including one to print the length of the \t{stats} array,
via this mechanism as the LLM constructs its response.  \systemname{}
echoes those commands and their outputs to the user.  Once the full
response has been processed, \systemname{} returns control to the
user.  As Section~\ref{sec:wheel} discusses, \systemname{}
augments the underlying debuggers with specialized commands for the LLM to use
when taking the wheel. For example, the \systemname{} variant for native code
installs debugger commands that utilize the \texttt{clangd}
language server~\cite{clangd,MicrosoftLSP} to retrieve source code
corresponding to symbol definitions.

\begin{figure}[pt!]
  \input{initialprompt}
  \caption{\label{fig:initial}\textbf{The initial prompt for the
  debugging session in Figure~\ref{fig:bootstrap} (\textsection\ref{sec:enriched}).} For brevity, the
  enriched stack includes only five lines of source in each frame, rather than the default of 10.}
  \Description{Text snippet of \systemname{}'s initial prompt.}
\end{figure}

\subsection{\label{sec:enriched} Initial Prompts and Enriched Stack Traces}

In addition to including the user's text, the initial prompt conveys
instructions to LLM and the context surrounding the
error.  We illustrate the components of the prompt in this
section, using the initial prompt in Figure~\ref{fig:initial} that
was generated for the first query in Figure~\ref{fig:bootstrap} as a running example.

\medskip
\noindent
\textbf{Instructions.}  The instructions at the top of the prompt ask
the LLM to answer questions about the root cause of the error, to
focus on user code, to explain values stored in variables, and to end
each response with either a fix or suggestions for further debugging
steps.  The last item ensures a relatively consistent structure for
answers that facilitates reading them and evaluating their quality.
Paragraphs 2-4 of the instructions are the \emph{take the wheel}
prompt described in Section~\ref{sec:wheel}.

\medskip
\noindent
\textbf{Enriched stack trace.}  \systemname{}'s success at identifying
and fixing errors relies critically on providing the LLM with
sufficient details to reveal the cause of the error.  A key source of
that information is the run-time stack.  Debuggers provide a
way for the user to view the stack trace but often only show function
names, source file locations, and possibly a couple lines of 
code for each stack frame.
\systemname{} provides a more detailed \emph{enriched stack
trace} to the LLM.  That stack trace includes the types and values
of variables for each frame, as well as a larger window of at least 10
lines of code.  Enriched stack traces also elide
frames corresponding to library code to better focus the LLM on
user-written code, which \systemname{} assumes to be the most likely
cause of errors.

In Python, \systemname{} leverages Pdb's internal data structures to
build enriched stack traces.  When converting values to suitable
string representations, \systemname{} must balance utility with the
size of the string produced.  
For objects, \systemname{} calls the object's
\inlinecode{__repr__} method if an appropriate (non-default) version
exists.  Otherwise, it iterates over the object's fields and recursively
converts their values to strings.
Similarly, \systemname{} recursively converts the values stored in
aggregate structures like lists, arrays, and dictionaries to strings,
but limits the number of elements shown to a small, fixed number.  The
rest of the elements are abbreviated with an ellipsis (...).
This recursive conversion of values to strings is limited to a depth
of three, at which point any remaining values are again abbreviated with
ellipses.  This strategy balances the need to provide the LLM with
sufficient information to diagnose the error with the need to avoid
overwhelming it with too much information.  In cases with the elided details
are important, the LLM can request them via the \emph{take the wheel}
mechanism.

\systemname{} follows roughly the same approach in LLDB, utilizing the
static types embedded in the DWARF debugging information to decode the
stack.  In addition, any pointers are dereferenced to show the values
being referred to as well; null pointers and illegal dereferences are dropped.

\smallskip
\noindent
\textbf{Inputs.}  The initial prompt also includes the
target's command line arguments and standard input when that
information is available from the underlying debugger.  These are
empty and elided in Figure~\ref{fig:initial}.

\smallskip
\noindent
\textbf{Error.}  A description of the error causing execution to stop
is extracted from the underlying debugger. When the error is due
to an assertion failure, \systemname{} instructs the LLM to assume
that the assertion is valid as written so that it will look beyond the
assertion for the real problem.

\smallskip
\noindent
\textbf{History.}
The initial prompt also includes the history of commands already
issued by the user, as well as their outputs.  This builds a more
complete context surrounding the user's query.

\subsection{\label{sec:wheel} Taking the Wheel}

\systemname{} supports \emph{take the wheel} debugging via the function call capabilities in OpenAI's API and most
recent models~\cite{OpenAIFunctions}.
This agentic approach~\cite{DBLP:journals/ker/WooldridgeJ95, DBLP:conf/iclr/GurFHSMEF24} lets clients
register callback functions with the LLM for obtaining additional
information while constructing a response.  The LLM calls these
functions by sending special messages to the client as part of its
response stream.  The client receives those messages, computes the
requested results, and sends them back to the LLM.  The initial prompt
describes how to use the available functions.

For example, \systemname{} registers a \t{debug(command)} function
for running a command in the underlying
debugger.  The LLM
calls \t{debug("p len(stats)")} through this mechanism in the session from Figure~\ref{fig:bootstrap}.
\systemname{} then runs Pdb's command
processing routine, \t{onecmd("p len(stats)")}, and captures the output to 
and send back.  \systemname{} similarly uses
the \texttt{SBCommandInterpreter.HandleCommand} routine in LLVM.  In both
cases, the command and output are printed so the user can see these
steps.

The LLM has sufficient background knowledge on debuggers and requires
\emph{no additional training} to navigate up/down the stack, inspect
variables and heap data, evaluate expressions, and perform other
typical debugger operations.

Supporting agentic reasoning over run-time program state via function calls is a
key technical innovation of \systemname{}.  Without this capability,
there would be no effective way to provide the LLM with a detailed view of relevant 
program state.  A common alternative technique for handling large amounts of
task-specific data in LLMs is to employ a retrieval augmented generation (RAG) model~\cite{DBLP:conf/nips/LewisPPPKGKLYR020}, which
collects and stores the data in a vectorized database that is then made available to the 
model for retrieval.  However, that approach seems less useful in this context, as program 
state information will be distinct for each debugging session and not easily vectorized.

\begin{table}[tp!]
  \centering\footnotesize
  \caption{\label{fig:extensions}\textbf{\systemname{} command
  extensions (\textsection\ref{sec:navigating-code}).}  \systemname{} extends the underlying debuggers
  with several new commands to help the LLM navigate through and
  understand the target's code.  
  \systemname{} provides access to them via the LLM's function call API.}
  \begin{tabular}{@{}lcp{3.5in}@{}}
    \toprule
    \multicolumn{1}{c}{\bf Command} &
    \multicolumn{1}{@{}c}{\bf Debugger} &
    \multicolumn{1}{c@{}}{\bf Output} \\
    \midrule
    \inlinecode{info symbol}       & Pdb  & The source code and/or docstring for a \t{symbol} referring to any function, method, field, class, or package. \\
    \texttt{slice symbol}          & Pdb  & The source code in the backwards slice of the global \t{symbol}. Interactive IPython/Notebook sessions only. \\      
    \texttt{code loc}              & LLDB & The source code surrounding \t{loc}, where \t{loc} has the form \t{filename:lineno}.\\
    \texttt{definition loc symbol} & LLDB & The declaration for the first occurrence of \t{symbol} at \t{loc}, where \t{loc} has the form \t{filename:lineno}.\\
    \bottomrule
  \end{tabular}
\end{table}

\subsection{Navigating the Code}
\label{sec:navigating-code}

While the LLM can often leverage pre-existing background knowledge of
common Python and C/C++ standard libraries, it will likely have
limited-to-no knowledge of any user-defined code or third-party
library functions. %
Trying to include all possibly-relevant source code in the initial
prompt would be infeasible and would prevent \systemname{} from scaling to
larger codebases.  Instead, \systemname{} 
extends the underlying debuggers with several new
commands designed to help the LLM navigate through and
understand the target's code.  These commands are available to the LLM
via function calls and listed in Table~\ref{fig:extensions}.

\systemname{} augments Pdb with the \t{info} command, which prints the
docstring for any function, class, field, method, or package.  It
additionally prints the source code for any user-defined function.
The \t{info} requests in Figure~\ref{fig:bootstrap} demonstrate these
two cases for \t{proportion\_blue} and \t{bootstrap\_statistic},
respectively.
The command is implemented via the standard
\t{inspect} and \t{pydoc} Python libraries.

The \t{info} command is not directly reproducible for unmanaged code
in LLVM because there is no comparable existing debugger support for
retrieving the source or documentation for a symbol.  Instead,
\systemname{} adds two other debugging commands to LLDB.  The first,
\t{code}, prints the code surrounding a source location described by a
filename and line number, as in \t{code polymorph.c:118}.
The second command, \t{definition}, prints the location and source code
for the definition corresponding to the first occurrence of a symbol on
a given line of code.  For example, \t{definition polymorph.c:118
target} prints the location and source for the declaration of
\t{target} corresponding to its use on that line.
The \t{definition} implementation leverages the \texttt{clangd}
language server, which supports source code queries via JSON-RPC and
Microsoft's Language Server Protocol~\cite{MicrosoftLSP}.

\subsection{\label{sec:slice-command} Slices for Interactive Python}

\systemname{} supports debugging interactive IPython sessions and
Jupyter notebooks.  Interactive sessions lead to many individual code
cells that are each evaluated separately.  Cells may be evaluated
out-of-order, override definitions from earlier cells, and
communicate values to other cells through top-level global variables.
Others have noted the challenges of reasoning about program behavior
in this
context~\cite{DBLP:conf/chi/HeadHBDD19,shankar2022bolt}.
\systemname{} provides an additional \t{slice} debugging command to
facilitate that reasoning.
The \t{slice} command computes the backwards slice for any variable
used in the current cell that was defined in previously-executed
cells.  It returns the code for cells in
that slice.
Suppose the code from
\t{bootstrap.py} in Figure~\ref{fig:bootstrap-src} were written in
four notebook cells as shown below:

\begin{flushleft}
\begin{minted}{python}
In[2]: def make_marble_sample(): ...

In[3]: def proportion_blue(sample): ...
    
In[4]: def resampled_stats(observed_marbles, num_trials):
           stats = bootstrap_statistic(observed_marbles,
                                       proportion_blue,
                                       num_trials)
           assert len(stats) == num_trials
           return stats
    
In[5]: observed_marbles = make_marble_sample()
       stats = resampled_stats(observed_marbles, 5)
\end{minted}
\end{flushleft}

\noindent
After evaluating these cells, \inlinecode{slice(observed_samples)}
returns the source for the cells labeled \t{In[2]}
and \t{In[5]}, and \inlinecode{slice(stats)} returns 
the source for all four cells.
\systemname{} uses \t{ipyflow} to
compute slices~\cite{shankar2022bolt,ipyflow}.

\subsection{Security and Risks}

It is possible for the LLM to issue debugging commands containing arbitrary code
through the \t{debug} function call provided by \systemname{}.  That code could,
for example, delete files or execute other malicious actions on the client.  
\systemname{} mitigates this risk by sanitizing LLM-generated debugging commands
before running them.  For Python, the sanitizer ensures that any functions
called in LLM-provided commands belong to a user-configurable whitelist.  For
native code, code provenance is harder to track and languages are more
permissive, so the sanitizer rejects any commands calling functions.  
\systemname{} supports an \inlinecode{--unsafe} flag to disable sanitizing when
the client system is running in an isolated environment that obviates the need
for such protections.

It is also possible for the LLM to hallucinate and respond with incorrect or
misleading diagnoses and fixes. \systemname{} mitigates this risk by not directly
applying proposed code fixes or suggestions to the target code. Instead, \systemname{} presents
them to the user, who may then vet and judge the quality of the LLM's
responses and decide whether or not to follow suggested changes.

\begin{table}
  \centering\footnotesize
  \rowcolors{3}{gray!15}{white} %

  \caption{\label{fig:python-benchmarks}\textbf{Python programs
  exhibiting a variety of common errors (\textsection\ref{sec:python}).}  Programs
  c1--c8 are command line scripts, and programs s1--s14 are Jupyter
  notebooks, which utilize
  two non-standard libraries consisting of 3,000 lines of
  code.  Semantic errors reflect failed tests expressed as
  assertions.  Crashes reflect unexpected termination due
  to any other type of error.}

  \begin{tabular}[ht]{lrccl}
    \toprule
    \multicolumn{1}{c}{\textbf{Name}} &
    \multicolumn{1}{c}{\textbf{LoC}} &
    \multicolumn{1}{c}{\textbf{Type}} &
    \multicolumn{1}{c}{\textbf{Reported Exception}} & 
    \multicolumn{1}{c}{\textbf{Root Cause}}\\
    \midrule
    c1 & 48 & semantic & Assertion Error & Off-by-one error in an h-index computation\\
    c2 & 81 & crash & Name Error  & Parameter not referenced properly\\
    c3 & 64 & crash & Value Error & Error in CSV column label leads to improper data parsing\\
    c4 & 89 & crash & Index Error & A class's \t{\_\_str\_\_} fails if an object's internal list is empty\\
    c5 & 29 & crash & Index Error & Missing one of two base cases in a recursive function\\
    c6 & 72 & crash & Name Error & Multiple errors related to building list of user-defined objects\\
    c7 & 71 & semantic & Assertion Error & Failure to convert input to lower case before processing\\
    c8 & 72 & semantic & Assertion Error & Missing test for lowercase words\\
    \midrule
    s1 & 123 & semantic & Assertion Error & Incorrect drop and rename operations leading to bad data\\
    s2 & 124 & semantic & Assertion Error & Incorrect max operation on a table\\
    s3 & 124 & semantic & Assertion Error & Incorrect aggregation function in pivot operation\\
    s4 & 124 & semantic & Assertion Error & Incorrect aggregation function in group operation\\
    s5 & 162 & semantic & Assertion Error & Hardcoded table data in wrong order\\
    s6 & 162 & crash & Name Error & Typo in variable reference\\
    s7 & 45 & semantic & Assertion Error & Function confuses parameter and global variable\\
    s8 & 49 & semantic & Assertion Error & Wrong percentile used in confidence interval construction\\
    s9 & 112 & semantic & Assertion Error & Wrong percentile used in confidence interval construction\\
    s10 & 118 & semantic & Assertion Error & Loops doesn't append to array correctly\\
    s11 & 181 & crash & Value Error & Creates a sample without replacement larger than the input\\
    s12 & 127 & crash & Value Error & Incorrect label when accessing column value for table row\\ 
    s13 & 127 & crash & Value Error & Pivot uses wrong columns for row/columns in new table\\
    s14 & 65 & crash & Index Error & Incorrect computation of random sample under null hypothesis\\
    \bottomrule
  \end{tabular}
\end{table}

\section{Evaluation}
\label{sec:evaluation}

We demonstrate \systemname{}'s capacity to identify the root cause of
defects and provide fixes in two contexts: bugs in relatively small
Python programs written by students and bugs in large C/C++ programs.
The former have well-defined expected behavior that enables us to
thoroughly and systematically assess \systemname{}.  The latter
demonstrates its effectiveness on unmanaged code when unusual corner
cases trigger crashes.
Our evaluation addresses the following research questions:
\textbf{RQ1:} Is \systemname{} effective at diagnosing and fixing bugs in Python?
\textbf{RQ2:} Which components of \systemname{} contribute to its effectiveness?
\textbf{RQ3:} Is \systemname{} effective at diagnosing and fixing bugs in unmanaged code (C/C++)?

\subsection{Python}
\label{sec:python}

We applied \systemname{} to all of the bugs in a collection of student labs from
two introductory computer science courses; see
Table~\ref{fig:python-benchmarks}.  Bugs c1--c8 are in non-interactive
scripts from a programming class that perform various file reading and
text processing tasks.  Bugs s1--s14 are in Jupyter
notebooks~\cite{jupyter-notebooks} from a data science class that
manipulate, visualize, and compute over arrays and tables.
Some bugs were apparent to the programs' authors.
Others were identified during autograding. %

Unlike many existing bug benchmarks for Python, these programs are
unpublished and thus not in the language model's training data.
In addition, the programs have clear correctness criteria that lead to
objective effectiveness metrics in our experiments.
The bugs are representative of common mistakes because they were introduced by real humans,
rather than synthetically generated.
They range from scoping issues, algorithmic errors, and
misuse of library functions to subtle misunderstanding of domain
knowledge.  They include both semantic errors leading to failed tests
and crashing errors that terminate execution abruptly.
Further, they reflect two important, widely-used modalities for Python
programming: non-interactive scripts and interactive computational
notebooks.  \systemname{} supports debugging in both settings.

Programs were prepared by removing them from their automatic grading
harness and replacing failed unit tests with \t{assert} statements
that generate exceptions.
We focus our evaluation on Python and perform an ablation study by
progressively enabling \systemname{} features.
We ran each program ten times under the five
configurations in Table~\ref{table:configurations}:
\config{Default Stack} includes standard stack traces, as generated by
\t{ipdb}~\cite{ipdb}, with 5 lines of code per frame in the initial
prompt, but it does not support the LLM taking the
wheel. \config{Enriched Stack} generates enriched stacks with ten lines
of code per frame, and \config{+Take the Wheel} additionally permits
\systemname{} to run debugger commands.  These three
configurations all use \t{why?} as the initial user text.
\config{+Targeted Question} asks a question specific to the failure.
For semantic errors, which validate the values stored in variables,
these questions describe what those values should be or what they
intend to represent.  For crashes, the questions relate the crash to
expected behavior, as in the following; we designed our questions to
be ``neutral'' and not hint at the root cause.
\begin{description}[labelindent=\parindent, itemindent=-\parindent]
  \item[\textbf{c3} (Crash)] Why am I not reading the CSV file
    correctly?
  \item[\textbf{s11} (Crash)] Why am I not able to sample 100 rows?
  \item[\textbf{c1} (Semantic)] Why am I not getting 3?
  \item[\textbf{s1}
      (Semantic)] \texttt{bill\_length\_mean\_by\_species} should be
    a table of the mean bill lengths of each species in our data
    set. Why isn't it?
\end{description}
The final \config{+Dialog} configuration is the same as \config{+Targeted
  Question} but extends the chat with a second query.  All
trials use the same follow up: \emph{Continue to explain your
  reasoning and give me a fix to make it work as I describe.}
Context-specific follow-ups work better, but we opted for
consistency.

\systemname{} used the \inlinecode{gpt-4-1106-preview} model for these
experiments.  Under \config{+Targeted Question}, the first prompt and
response led to, on average, a chat of about 10,000 tokens (7,500
words), a cost of about \$$0.12$ USD under OpenAI's current pricing
model~\cite{OpenAIPricing}, and a completion time of about 25
seconds.  Subsequent steps in extended debugging dialogs incurred
comparable costs.  Time was highly variable and dominated by the
performance of OpenAI's service.  These characteristics will be
different for other platforms and models and, given current trends, we expect
significant reductions in both time and cost as models improve.

\begin{table}[tp!]
  \centering\footnotesize
  
  \caption{\label{table:configurations}\textbf{Configurations used in the Python experiments (\textsection\ref{sec:python}).}}

  \begin{tabular}{lccccc}
  \toprule
  \multirow{2}{*}{\textbf{\centering Configuration}} 
    & {\textbf{Stack}} 
    & {\textbf{Take the}} 
    & {\textbf{Initial}} 
    & {\textbf{Ask a}} \\%
    & {\textbf{Trace}} 
    & {\textbf{Wheel}} 
    & {\textbf{Prompt}} 
    & {\textbf{Follow-up}} \\%
  \midrule
  {Default Stack} & standard & & \texttt{why?} \\%
  {Enriched Stack} & enriched & & \texttt{why?} \\%
  {+Take the Wheel} & enriched & \checkmark & \texttt{why?} \\%
  {+Targeted Question} & enriched & \checkmark & \emph{specialized} \\%
  {+Dialog} &  enriched & \checkmark & \emph{specialized} & \checkmark \\%
  \bottomrule
  \end{tabular}
  \end{table}

  \begin{figure}[tp!]
    \includegraphics[width=0.65\columnwidth]{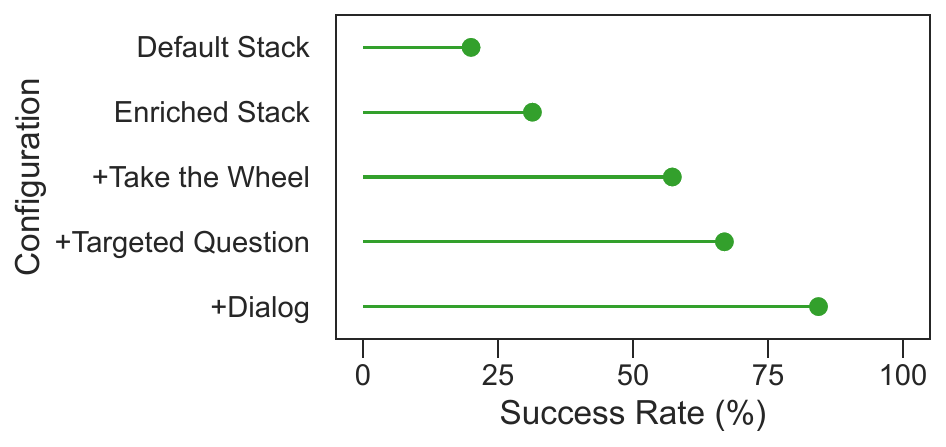}
    \caption{\label{fig:summary}\textbf{Overall \systemname{} success rate for each configuration (\textsection\ref{sec:python}).}
    \systemname{} innovations and user-provided context gradually increase effectiveness.}
    \Description{Chart showing the overall success rate of \systemname{} under each configuration.
      Each feature addition improves the success rate, culminating at 85\% using the fully-featured \systemname{}.
    }
  \end{figure}

\smallskip
\noindent
\textbf{RQ1: Is \systemname{} effective at diagnosing and fixing bugs in Python?}

Each response was manually examined and deemed a success if it
included an accurate explanation of the error and an actionable fix.
That fix could be either code or a prose description in which all
necessary details were made explicit.  To avoid bias in this assessment,
explicit criteria for each program was determined prior to examining
the responses.

Figure~\ref{fig:summary} shows the success rate under each
configuration.  The simplest configuration, \config{Default Stack},
provides functionality roughly equivalent to the user copying and
pasting the program stack trace and basic error information into an
LLM chat window and requesting a fix. We use this configuration as a
baseline for evaluating the impact of \systemname{}'s more advanced
configurations. With all features enabled, \systemname{} was successful at identifying
and fixing bugs in well over half of the trials.  Any time or energy
expended by the user manually debugging those cases would be all but
eliminated by using \systemname{}.

\smallskip
\begin{tcolorbox}[colback=green!5!white,colframe=green!75!black, arc=1mm, top=1mm, bottom=1mm, left=1mm, right=1mm]
\textbf{RQ1 Summary:}  Even with just the simple question \t{why?},
\systemname{} was successful 57\% of the time.  With
questions specialized to the target's particular error, that number
jumps to 67\%, and with an additional dialog step \systemname{}
succeeded in identifying and fixing the defect in 85\% of the trials.
\end{tcolorbox}

\smallskip
\noindent
\textbf{RQ2: Which components of \systemname{} contribute
  to its effectiveness?}

\begin{figure}[tp!]
  \begin{center}
  \includegraphics[width=\textwidth]{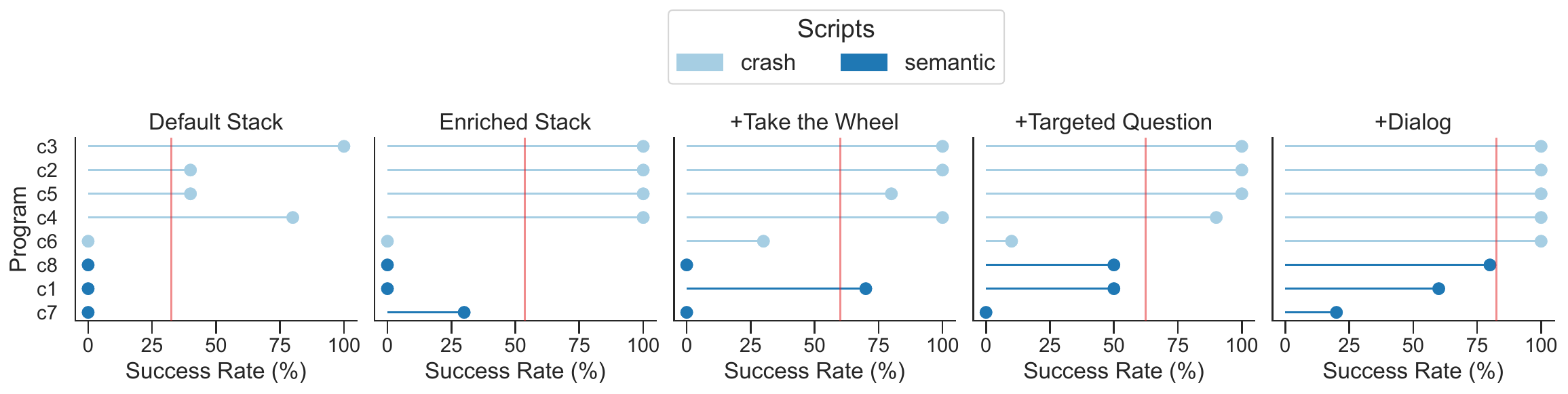} \\%
  \includegraphics[width=\textwidth]{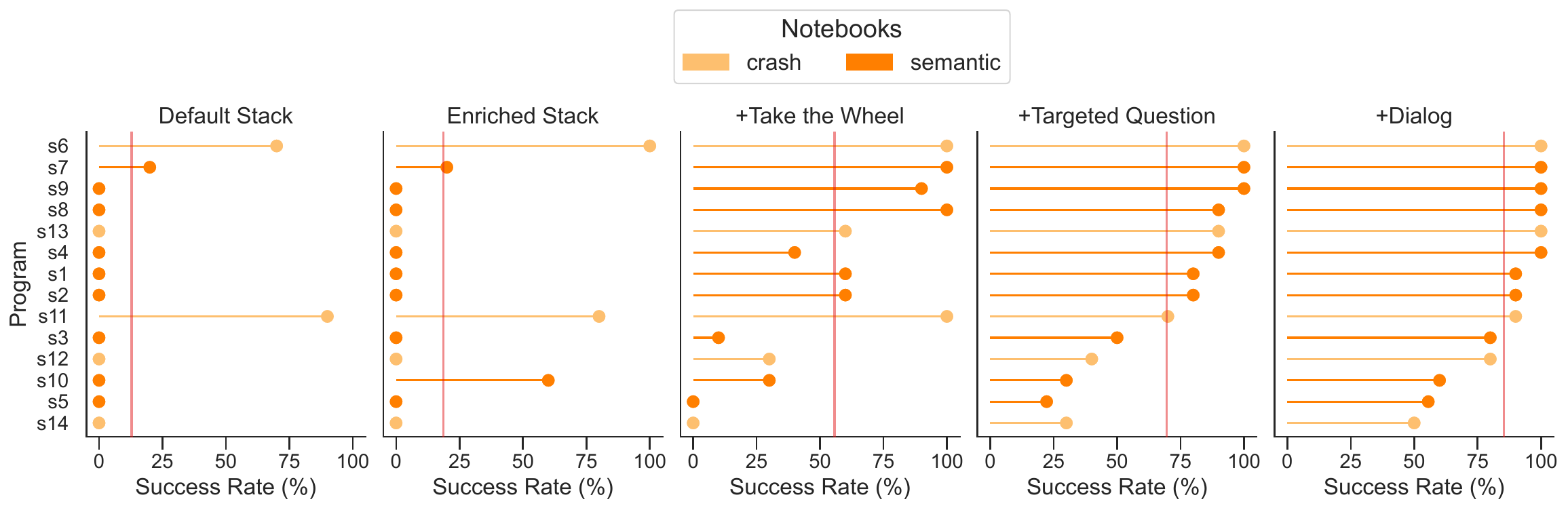}
  \end{center}
  \caption{\label{fig:quality}\textbf{Success rate for \systemname{}
      for each program and configuration (\textsection\ref{sec:python}).}  Vertical lines show the mean.}
  \Description{A detailed version of Figure~\ref{fig:summary}, showing the success rate for each program evaluated under each
  configuration. While there is some variation, the \systemname{} generally improves with additional features.}
\end{figure}

Figure~\ref{fig:quality} presents the success rates for each program
under each configuration.  The \config{Enriched Stack} plots
demonstrate that enriched stacks provide some benefit, particularly for
crashes in which the stack contains sufficient information to diagnose
the problem, but they alone do not provide much improvement
for many semantic errors in which the relevant computation steps
complete before failure.  However, enriched stacks coupled with
letting the LLM take the wheel led to significant improvement in the
success rate for both crashing and semantic bugs, as shown in the
\config{+Take the Wheel} plots.

Using the \config{+Take the Wheel} feature, the LLM issues
from 0 to 12 debugging commands per run, most commonly calling
the \t{info}, \t{slice} (for notebooks), and
\t{p} (print) debugging commands.  While all of these commands provide useful information about execution state and code,
the \t{slice} command was critically
important for notebooks.  Without it, success rates rarely
improved when the LLM took the wheel.

The \config{+Targeted Question} configuration demonstrates the impact
of providing even the most modest details about expected behavior in queries.  
When the LLM is asked to continue its reasoning in \config{+Dialog},
\systemname{}'s success rate improves despite the follow-up prompt
providing no feedback on the contents or quality of the first
response.  This phenom indicates that constraints on the underlying LLM's
response lengths may prevent it from conducting the amount of
reasoning necessary to develop a fix in a
single step.
The success rates for \config{+Targeted Question} and \config{+Dialog}
demonstrate the importance of continued dialogs and user input.  We expect
those features to be even more important to \systemname{}'s
success when diagnosing bugs in more complex programs.

The LLM also demonstrated its background knowledge with the responses
including, for example, details of Python idioms and libraries, the
definition of h-index~\cite{DBLP:journals/pnas/Hirsch05}, and the
implementation and limitations of various statistical techniques.

Failures were generally due to the LLM not always recognizing or discovering key
aspects of a program's behavior.
We observe that in some cases, enabling additional features in \systemname{} decreases
its success rate. We attribute this result to the fact that longer and more complex prompts
can occasionally degrade the effectiveness of LLMs~\cite{DBLP:conf/acl/LevyJG24}.
In general, the further the distance between the root cause of a bug
and observable effect, the more challenging it was for \systemname{}
(and people~\cite[p. 243]{ZellerDebugging}) to find it.
In some cases, it
was on the right track but did not converge on an actionable fix.  In
others, it suggested changes that would introduce other bugs.  It also
occasionally made mistakes, such as conflating proportions and
percentages or failing to handle unusual corner cases.
All of these
could be mitigated by feedback from the user in subsequent follow-ups.

\smallskip
\begin{tcolorbox}[colback=green!5!white,colframe=green!75!black, arc=1mm, top=1mm, bottom=1mm, left=1mm, right=1mm]
\textbf{RQ2 Summary:} While all features of \systemname{} contribute to
its success, the technical innovations enabling it to take the wheel
are critical.  The most sophisticated configurations show that
user-provided contextual information about behavior and engaging in
multi-step dialogs are particularly good ways to improve its
effectiveness.
\end{tcolorbox}

\subsection{C and C++}
\label{sec:native}

Programs in unmanaged languages such as C and C++ are vulnerable to
memory safety errors.  These memory errors can also hinder the
debugging process: the crash may not occur immediately at the memory
violation but instead much later on, and the crash may cause
corruption of the stack and/or heap, making it challenging to recover
any useful information.

Table~\ref{table:c-cpp-programs} summarizes the programs extracted from the BugBench~\cite{lu2005bugbench}
and BugsC++~\cite{DBLP:conf/kbse/AnKCYY23} suites
used to evaluate \systemname{}'s effectiveness at debugging unmanaged code.
Programs used in this evaluation are all real-world applications with concrete known bugs.
The four BugBench programs were selected as the only ones we could retrieve, build, and reproduce on our system.
The BugsC++ suite does not include the original crash-causing inputs.
However, it provides links to the original bug report, CVE identifier, and/or exploit-fixing patch, from which
we manually retrieve crash reproduction information. We randomly selected and reproduced four bugs
from the ``memory error'' category.

\begin{table}[tp]
  \centering\footnotesize
  \rowcolors{3}{white}{gray!15} %
  
  \caption{\label{table:c-cpp-programs}\textbf{Bugs in unmanaged C/C++
      code, and our criteria for fixing the proximate
      cause or the root cause of each (\textsection\ref{sec:native}).}}
  \begin{tabular}{llp{0.5in}p{1.7in}|>{\raggedright\arraybackslash}p{0.8in}@{}>{\raggedright\arraybackslash}p{0.8in}}
  \toprule
   &
   &
 \multicolumn{1}{c}{\textbf{Error}} &
 \multicolumn{1}{c}{\textbf{}} &
 \multicolumn{2}{|c}{\textbf{Fix}} 
 \\%
  \multicolumn{1}{c}{\textbf{Program}} &
  \multicolumn{1}{c}{\textbf{LoC}} &
 \multicolumn{1}{c}{\textbf{Type}} &
 \multicolumn{1}{c}{\textbf{Root Cause}} &
 \multicolumn{1}{|c}{\textbf{Proximate Cause}} &
  \multicolumn{1}{c}{\textbf{Root Cause}}
 \\\midrule%
  BC~\cite{lu2005bugbench}              & 17.0k & Buffer overflow &  Input from data file printed to a fixed-size buffer                          & Truncate on copy         & Use dynamic size \\%
  GZIP~\cite{lu2005bugbench}            & 8.2k  & Buffer overflow & Command line argument unsafely copied to a fixed-size buffer                  & Truncate on copy         & Check size \& warn/exit \\%
  NCOM~\cite{lu2005bugbench}            & 1.9k  & Buffer overflow & Command line argument unsafely copied to a fixed-size buffer                  & Truncate on copy         & Check size \& warn/exit \\%
  PEG~\cite{DBLP:conf/kbse/AnKCYY23}    & 14.7k & Null dereference &  Invalid input produces corrupted data structure                              & Check if not null        & Warn/exit \\%
  POLY~\cite{lu2005bugbench}            & 0.7k  & Buffer overflow & Command line argument is unsafely copied to a fixed-size buffer               & Truncate on copy         & Check size \& warn/exit \\%
  TIFF~\cite{DBLP:conf/kbse/AnKCYY23}   & 58.9k & Division by zero & Combination of command line options leads to a division by zero             & Override option values   & Warn/exit when invalid \\%
  YAML1~\cite{DBLP:conf/kbse/AnKCYY23}  & 8.7k  & Stack overflow   & Long sequences of \texttt{\{} in the input leads to deep recursion & Use iterative method     & Guard recursion depth \\%
  YAML2~\cite{DBLP:conf/kbse/AnKCYY23}  & 8.7k  & Assertion failure & Specific input causes a peek request for non-existent ``next'' token        & Replace assert           & Check before peeking \\%
  \bottomrule
  \end{tabular}
\end{table}

Some of the programs studied do not crash at run time.  We employed
AddressSanitizer~\cite{DBLP:conf/usenix/SerebryanyBPV12} to %
force a crash at the moment a memory violation occurs to trigger
those defects.
AddressSanitizer is already capable of reporting some information
about the crash when it happens. However, this information is often
very dense, and typically points at the symptom of the bug, not its
root cause. We did not include that information in the initial prompt.%

\smallskip
\noindent
\textbf{RQ3: Is \systemname{} effective at diagnosing and fixing bugs in unmanaged code (C/C++)?}

We ran our C/C++ experiments on an x86 Ubuntu 22.04 server.  We used
Clang and LLDB 17 to compile and debug, using flags \texttt{-g -Og
  -fno-omit-frame-pointer}. \systemname{} used OpenAI's
\texttt{gpt-4-1106-preview} model.  Each program was run ten times
using queries of the form \texttt{I am debugging cpp-peglib.  Provide
  the root cause of this crash}, for PEG, followed by a request to
include code in the response.  Average time (27 seconds) and cost
(\$$0.06$ USD) were comparable to Python. %

We manually examined each response to determine if \systemname{}
successfully provided an actionable code fix for the proximate cause
of the crash or for the underlying root cause.  We used the criteria
outlined in Table~\ref{table:c-cpp-programs}.
While fixing root causes is the ultimate goal, fixing proximate
causes can still be beneficial as fixing crashes enables further debugging steps.

\begin{figure}[tp]
  \includegraphics[width=0.75\columnwidth]{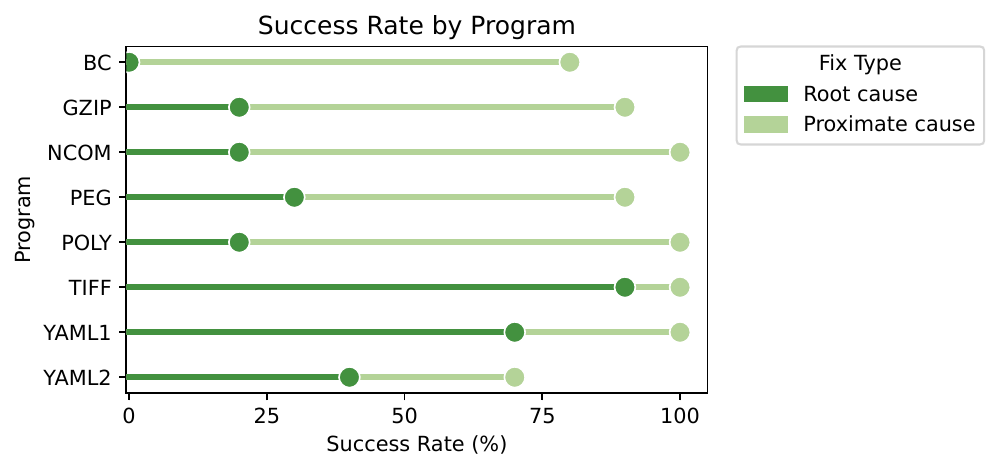}
  \caption{\label{fig:c-success}\textbf{\systemname{} success rate at
      fixing the proximate or root cause in C/C++ programs (\textsection\ref{sec:native}).}
    \systemname{} successfully identified and fixed the root cause
    36\% of the time and the proximate cause an additional 55\% of the
    time.} %
    \Description{Chart showing the success rate of \systemname{} at fixing the proximate or root cause in C/C++ programs.
    Proximate causes are identified and fixed most of the time, while success finding and fixing root cause varies
    among programs.}
\end{figure}

Figure~\ref{fig:c-success} presents \systemname{}'s ability to suggest a fix
for either the proximate or root cause of the bug.
Generally, \systemname{} is excellent at diagnosing and explaining the reason for the crash,
which in itself may be useful to programmers.
For BC, GZIP, NCOM, and POLY, \systemname{} tends to suggest replacing
the \texttt{strcpy} or \texttt{sprintf} call with their respective
\texttt{strncpy} and \texttt{snprintf} counterparts to prevent buffer
overflows.  While correct, this change truncates the input silently.
Validation or other measures should be added to obtain a robust
fix. The root cause in BC is inside code generated from a YACC file.
The \texttt{clangd} language server does not handle this case in a way
that would let \systemname{} answer the LLM's \t{definition} requests properly.

In the case of PEG, \systemname{} correctly identifies which pointer
is null but typically suggests ignoring it instead of failing
immediately. This is similar to YAML2, where \systemname{} recommends
replacing the assertion with a check inside a function rather than
recommending that the client check that the function's preconditions
are met prior to the call.
\systemname{} has a relatively high root cause fix rate for YAML1 and
TIFF.  It often correctly suggests fixes to limit recursion depth
(YAML1) and to validate input parameters (TIFF).

\smallskip
\begin{tcolorbox}[colback=green!5!white,colframe=green!75!black, arc=1mm, top=1mm, bottom=1mm, left=1mm, right=1mm]
\textbf{RQ3 Summary:}
\systemname{} was successful in virtually all of our trials in diagnosing and
explaining the cause of the crash. It was also capable of providing relevant,
actionable fixes: 36\% of its suggestions addressed the root cause of the bug,
while another 55\% corrected the proximate cause. %
\end{tcolorbox}

\subsection{Threats to Validity}
\label{sec:threats}

This paper evaluates \systemname{} on two suites of code. The primary
suite is a collection of unpublished student labs that may not be
entirely representative of code written by, for example, experienced
programmers.  The second suite consists of real C/C++ applications and
bugs drawn from the BugBench and BugsC++ suites.  Unlike the Python
suite, the C/C++ source code and the bug fixes for these programs are
available on GitHub, which may lead to data leakage affecting the
C/C++ study if those repositories were part of the training set for
the LLMs we used.  While the C/C++ suite consists of real-world
applications, most of the errors are memory errors.  Other types, such as assertion failures, concurrency errors, or other
logical errors, may lead to different results.

\systemname{} depends on an LLM to analyze and drive exploration of
state, and like all systems based on LLMs today, its performance is
affected by prompt engineering. It is possible that \systemname{}'s
prompts are overfit to the specific GPT-4 models we employed; this
threat is somewhat mitigated by the fact that \systemname{} was
originally developed using a different model (GPT-3.5-turbo). LLMs
are also inherently stochastic, and it is possible to obtain unusually
good results by chance. To mitigate this threat, our evaluation runs
\systemname{} on each test program at least ten times, which produced
stable and repeatable results with only small variation in aggregate.

Our evaluation depends on a manual evaluation of
whether \systemname{}'s explanation of a bug and its proposed fix are
satisfactory.  We mitigated the risks of subjectivity by
using precisely defined criteria decided upon in advance.  Python
fixes were deemed successful if the resulting code met the correctness
requirements outlined in the assignment.  C/C++ fixes were deemed
successful at fixing the proximate or root cause using the criteria in
Table~\ref{table:c-cpp-programs}.  Fixes described in prose were
permitted, provided that the details of all necessary changes to the code were made
explicit.

\section{Future Work}
\label{sec:futurework}

We see several promising avenues for future work. Incorporating existing
fault localization approaches into \systemname{}, rather than relying
solely on the LLM's ability to explore the program's code and
state, could potentially increase its effectiveness and efficiency by
allowing the LLM to focus its attention on suspicious files, functions, or lines of source code. Similarly,
incorporating delta debugging~\cite{DBLP:conf/esec/Zeller99} could
increase the effectiveness of \systemname{} by limiting the amount of
input for an LLM and providing failure-inducing events as
guidance. Finally, integrating \systemname{} with a time-travel
debugger~\cite{gdb-reverse-debugging,DBLP:conf/usenix/OCallahanJFHNP17} would expand its reach to exploring program state over time,
letting it answer queries that cannot be answered given only the current
program state.  One challenge of integrating these more sophisticated techniques
will be ensuring that the LLM can effectively utilize them, which may 
necessitate fine tuning or additional training on their usage.

\section{Conclusion}
\label{sec:conclusion}

This paper presents \systemname{}, the first AI-based debugging
assistant. Our evaluation 
shows that
engaging in a debugging dialog with \systemname{} can significantly
assist in identifying root causes of errors and developing correct
fixes.

\begin{acks}
This material is based upon work supported by the National Science Foundation
under Grant No.~2243636.  Any opinions, findings, and conclusions
or recommendations expressed in this material are those of the author(s)
and do not necessarily reflect the views of the National Science Foundation.
\end{acks}

\section*{Data-Availability Statement}
\systemname{} is available on GitHub at
\texttt{\href{https://github.com/plasma-umass/ChatDBG}{github.com/plasma-umass/ChatDBG}}.
An archived version is also available on Zenodo~\cite{levin_2025_15185773}.

\bibliographystyle{ACM-Reference-Format}
\bibliography{references}{}
\end{document}